\documentstyle[12pt]{article}

\def\be{\begin{equation}}
\def\ee{\end{equation}}
\def\ba{\begin{array}{c}}
\def\baa{\begin{array}{ll}}
\def\ea{\end{array}}

\def\ben{$$}
\def\een{$$}
 
\begin{document}

\titlepage

  \begin{center}{\Large \bf
Discrete ${\cal PT}-$symmetric square-well oscillators
 }\end{center}

\vspace{5mm}

  \begin{center}

Miloslav Znojil\footnote{ e-mail: znojil@ujf.cas.cz}

 \vspace{3mm}

\'{U}stav jadern\'e fyziky AV \v{C}R, 250 68 \v{R}e\v{z}, Czech
Republic\\

 \vspace{3mm}

\end{center}

\vspace{5mm}


\section*{Abstract}

Exact solvability of the discretized $N-$point version of the
${\cal PT}-$symmetric square-well model at all $N$ is pointed out.
Its wave functions are found proportional to the classical
Tshebyshev polynomials $U_k\cos \theta)$ of a complex argument. A
compact secular equation is derived giving the real spectrum of
energies at all the non-Hermiticity strengths $Z \in (-
Z_{crit}(N), Z_{crit}(N))$. In the limit $Z \to 0$ the model
degenerates to a Hermitian H\"{u}ckel Hamiltonian.

 \vspace{9mm}

\noindent PACS  03.65.Ge, 02.60.Lj,  02.70.Bf, 31.15.Ct


\vspace{9mm}


 \newpage

\section{Introduction}

The essence of the concept of the so called ${\cal PT}-$symmetric
Quantum Mechanics \cite{BB} lies in a half-forgotten fact that
although the operators of observables (say, Hamiltonians $H$) {\em
must} be Hermitian ({with respect to {a} metric} $\Theta$  in
Hilbert space), they {are {\em also} allowed to} be Hermitian
{with respect to a {\em nontrivial} metric},
 \be
 H^\dagger = \Theta\,H\,\Theta^{-1}, \ \ \ \ \ \ \ I \neq \Theta
 = \Theta^\dagger >0.
 \label{quasiher}
 \ee
In sufficient detail, the formal aspects of this idea were already
well described in the review paper \cite{Geyer} showing that in
nuclear physics a decisive {simplification} of certain
Schr\"{o}dinger equations may be achieved after a formal
transition from the most common $\Theta_1=I$ to another $\Theta_2
\neq I$.

Beyond the area of nuclear physics the {feasibility} of the
necessary calculations represents a key technical challenge for
$\Theta \neq I$. Fortunately, for some differential Hamiltonians
$H= \hat{p}^2+V(x)$, an unexpectedly satisfactory answer has been
found in a factorization of $\Theta$, typically, into a product of
parity ${\cal P}$ and the so called quasi-parity~${\cal
Q}$~\cite{pseudo} or charge ${\cal C}$~\cite{BBJ}. In the other
words, one requires that the observables are {\em both} ${\cal
PT}-$symmetric and ${\cal CP}-$pseudo-Hermitian. This means that
all our non-Hermitian observables must be compatible with
eq.~(\ref{quasiher}) {\em and} that they must also commute with
certain operator ${\cal PT}$. In the terminology advocated by A.
Mostafazadeh \cite{ali} the latter requirement should be
generalized and re-interpreted as the so called ${\cal
P}-$pseudo-Hermiticity relation
 \be
 H^\dagger = {\cal P}\,H\,{\cal P}^{-1}, \ \ \ \ \ \ \ I \neq {\cal P}
 ={\cal P}^\dagger
 \label{pseudoher}
 \ee
In such a widely accepted scenario \cite{lite} our observables may
remain manifestly non-Hermitian in the current sense, $H \neq
H^\dagger$. Still, their spectra {\em must} remain real and the
work with the underlying indefinite `pseudo-metric' operators of
generalized parity ${\cal P}$ {\em should} remain sufficiently
easy.

In the light of these two requirements we feel particularly
inspired here

\begin{itemize}

 \item {\bf [a]}
by the  rigorous Krein-space analysis~\cite{Langer} of the reality
of spectra of quantum particles confined inside a one-dimensional
${\cal PT}-$symmetric box $V(x)$,

 \item {\bf [b]}
in context~{\bf [a]}, by our older explicit
construction~\cite{sqw} of bound states in one of the most
elementary square-well forms of $V(x)$,

 \item {\bf [c]}
in context~{\bf [b]}, by the very recent Weigert's~\cite{Weigert}
three-by-three matrix discretization
 \be
 \left (
 \begin{array}{ccc}
 2+\frac{1}{4}\,{\rm i}\,{Z}&-1&0\\
 -1&\ \,2&-1\\
 0&-1&2-\frac{1}{4}\,{\rm i}\,{Z}
 \ea
 \right )\,
 \left (
 \ba
 \alpha_0\\
  \gamma\\
 \beta_0
 \ea
 \right )
 =
 \frac{1}{4}\,E
 \,
 \left (
 \ba
 \alpha_0\\
  \gamma\\
 \beta_0
 \ea
 \right ).
 \label{matr4}
 \ee
of the oscillator of ref.~\cite{sqw}.
\end{itemize}
 \noindent
In this general framework we intend to start from the ${\cal
PT}-$symmetric ordinary differential Schr\"{o}dinger equation
 \be
 \left [ -\frac{d^2}{dx^2} + V(x)
 \right ]\,
 \psi(x)=E\,\psi(x), \ \ \ \ \ \ \ \ \
 \psi(\pm 1)=0
 \,, \ \ \ \  \ \ \ \ \ \
 V(x) = \left [ V(-x)
 \right ]^*
 \label{basic}
 \ee
and from the proof given in ref.~\cite{Langer} that the resulting
spectrum of energies $E=E_n$, $n=0, 1, \ldots$ is {\em real} and
discrete for {\em all} the {complex} ${\cal PT}-$symmetric
potentials $V(x)$ which are not too strong,
 \be
 \|V\|_\infty <\frac{3}{8}\,\pi^2 \approx 3.701\,.
 \label{potcy}
 \ee
In the spirit of item {\bf [b]} we shall only pay attention to one
of the simplest piecewise constant and purely imaginary
antisymmetric potentials
 \be
 V(x) =
 \left \{
 \begin{array}{ll}
 +{\rm i}\,Z & x\in (-1,0),
 \\[0.1cm]
 -{\rm i}\,Z & x\in (0,1 ).
 \ea
 \right .
 \label{potenc}
 \ee
Along the lines of item {\bf [c]} we shall introduce the
Runge-Kutta discrete lattice of coordinates,
 \ben
 x_0=-1, \ \ \ \ \ \ \ \ x_k=x_{k-1}+h=-1+kh, \ \ \ \ \ \ \ \
  h = \frac{2}{N}, \ \ \ \ \ \ \ \
  k = 1, 2, \ldots, N\,
 \een
and pay attention to the general discrete analogue
 \be
  -
 \frac{\psi(x_{k+1})-2\,\psi(x_{k})+\psi(x_{k-1})}{h^2}
  -{\rm i}\,{\rm sign}(x_k)
 Z\,\psi(x_{k})
 =E\,\psi(x_k)
 \label{diskr}
 \ee
of our differential Schr\"{o}dinger equation (\ref{basic}) +
(\ref{potenc}). In combination with the boundary conditions
 \ben
  \psi(x_0)=\psi(x_N)=0,
  \een
this in fact represents a generalization of the Weigert's  $N=4$
eq.~(\ref{matr4}) to all the integers $N$.

\section{Models with the even $N = 2n+4$ \label{above} }

Let us recollect that the continuous $N = \infty$ model
(\ref{potenc}) is exactly solvable \cite{sqw}. The solvability in
closed form also characterizes its modifications with periodic
boundary conditions and/or more discontinuities~\cite{triper}. In
this context, the exact solvability of the following
$(N-1)-$dimensional matrix generalization
 \be
 \left (
 \begin{array}{cccc|c|ccc}
 {\rm i}\xi-F&-1&&&&&&\\
 -1&{\rm i}\xi-F&\ddots&&&&&\\
 &\ddots&\ddots&-1&&&&\\
 &&-1&{\rm i}\xi-F&-1&&&\\
 \hline
 &&&-1&-F&-1&&\\
 \hline
 &&&&-1&-{\rm i}\xi-F&\ddots&\\
 &&&&&\ddots&\ddots&-1\\
 &&&&&&-1&-{\rm i}\xi-F
 \ea
 \right )\,
 \left (
 \ba
 \alpha_0\\
 \alpha_1\\
 \vdots\\
 \alpha_n\\
 \hline
 \gamma\\
 \hline
 \beta_n\\
 \vdots\\
 \beta_0
 \ea
 \right )
 =0
 \label{matr}
 \ee
of the three-dimensional Weigert's model~(\ref{matr4}) with the
re-scaled energy eigenvalues $F=E\,h^2-2$ and with the re-scaled
strength $\xi=Z\,h^2$ of the non-Hermiticity would not be too
surprising.

In the first step of our analysis, due to the ${\cal PT}-$symmetry
of our problem we may set
 \ben
 \alpha_k=a_k+{\rm i}\,b_k, \ \ \ \ \ \ \
 \beta_k=a_k-{\rm i}\,b_k\equiv \alpha_k^*, \ \ \ \ \ \ \ \
 k = 0, 1, \ldots, n
 \een
with some {\em real}  elements $\gamma=\psi(0)$, $a_k= {\rm
Re}\,\psi\left ( x_{k+1}\right )$ and $b_k= {\rm Im}\,\psi\left (
x_{k+1}\right )$. Once we recollect the definition of the
classical Tshebyshev polynomials of the second kind \cite{Ryzhik},
 \ben
 U_k(\cos \theta) = \frac{\sin (k+1)\theta}{\sin \theta},
 \ \ \ \ \ \ k = 0, 1, \ldots,
 \een
we immediately obtain the wave functions in closed form,
 \be
 \alpha_k
 =U_k
 \left (
 \frac{-F+{\rm i}\xi}{2}
 \right )\, (
 a+{\rm i}b ),
 \ \ \ \ \ \ \ \ \ k=0,1,\ldots, n\,.
 \label{eigenvecs}
 \ee
This reduces the full tridiagonal $(N-1) \times (N-1)-$dimensional
matrix eq.~(\ref{matr}) to the mere three matching conditions,
 \be
 \left (
 \begin{array}{ccccc}
 -1&{\rm i}\xi-F&-1&0&0\\
 0&-1&-F&-1&0\\
 0&0&-1&-{\rm i}\xi-F&-1
 \ea
 \right )\,
 \left [
 \ba
 U_{n-1}
 \left (
 \frac{-F+{\rm i}\xi}{2}
 \right )\,
 \left (
 a_{}+{\rm i}b_{}\right )\\
 U_n
 \left (
 \frac{-F+{\rm i}\xi}{2}
 \right )\,
 \left (
 a_{}+{\rm i}b_{}\right )\\
 \gamma\\
 U_{n}
 \left (
 \frac{-F-{\rm i}\xi}{2}
 \right )\,
 \left (
 a_{}-{\rm i}b_{}\right )\\
 U_{n-1}
 \left (
 \frac{-F-{\rm i}\xi}{2}
 \right )\,
 \left (
 a_{}-{\rm i}b_{}\right )
 \ea
 \right ]
 =0.
 \label{redmatr}
 \ee
The first and the third lines may be simplified to give
 \be
 \gamma= U_{n+1}
 \left (
 \frac{-F+{\rm i}\xi}{2}
 \right )\,
 \left (
 a_{}+{\rm i}b_{}\right )
 =
 U_{n+1}
 \left (
 \frac{-F-{\rm i}\xi}{2}
 \right )\,
 \left (
 a_{}-{\rm i}b_{}\right ).
 \label{redmatrbe}
 \ee
The middle line defines the product
 \be
 F\,\gamma= -U_{n}
 \left (
 \frac{-F+{\rm i}\xi}{2}
 \right )\,
 \left (
 a_{}+{\rm i}b_{}\right )
 -
 U_{n}
 \left (
 \frac{-F-{\rm i}\xi}{2}
 \right )\,
 \left (
 a_{}-{\rm i}b_{}\right ).
 \label{redmatral}
 \ee
This forces us to separate the $F=0$ case as exceptional.

\subsection{The existence of a nontrivial solution at $F=0$ }

Once we set, tentatively, $F=0$, it is easy to deduce from
eq.~(\ref{redmatrbe}) that the parameter $ a_{}$ must vanish for
the even $n = 0, 2, 4, \ldots$ (and we may normalize $b=1$) while
$ b_{}=0$ and $a=1$ for the odd $ n = 1, 3, 5, \ldots$. Thus,
eq.~(\ref{redmatrbe}) degenerates to the mere definition of the
last element $\gamma$ of the eigenvector and we are left with the
single secular eq.~(\ref{redmatral}) which acquires the following
two alternative forms,
 \be
 \ba
 U_{n}
 \left (
 \frac{1}{2}\,{\rm i}\,\xi
 \right )-
 U_{n}
 \left (
 \frac{1}{2}\,{\rm i}\,\xi
 \right )=0,
 \ \ \ \ n = 2m,
 \\
 U_{n}
 \left (
 \frac{1}{2}\,{\rm i}\,\xi
 \right )
 +
 U_{n}
 \left (
-\frac{1}{2}\,{\rm i}\,\xi
 \right )=0,
 \ \ \ \ n = 2m+1.
 \ea
 \label{redmatralbe}
 \ee
At any $m = 0, 1, \ldots$ these conditions are satisfied
identically. We may conclude that our tentative ``guess of the
energy" was correct and that $F=0$ is always the eigenvalue. It is
remarkable that in spite of the non-Hermiticity of the
Hamiltonian, this  ``robust" eigenvalue remains real at {\em all}
the real couplings $Z \in (-\infty,\infty)$.

\section{Closed secular equations for $N = 2n+4$ \label{rabove} }


Whenever $F \neq 0$ we may treat eq.~(\ref{redmatrbe}) not only as
the condition of vanishing of the imaginary part of $\gamma$,
 \be
  U_{n+1}
 \left (
 \frac{-F+{\rm i}\xi}{2}
 \right )\,
 \left (
 a_{}+{\rm i}b_{}\right )
 =
 U_{n+1}
 \left (
 \frac{-F-{\rm i}\xi}{2}
 \right )\,
 \left (
 a_{}-{\rm i}b_{}\right )
 \label{redmatrbex}
 \ee
but also as an explicit definition of the non-vanishing
left-hand-side quantity $F\gamma$ in eq.~(\ref{redmatral}). Its
insertion simplifies the latter relation,
 \be
 T_{n+1}
 \left (
 \frac{-F+{\rm i}\xi}{2}
 \right )\,
 \left (
 a_{}+{\rm i}b_{}\right )
 =-
 T_{n+1}
 \left (
 \frac{-F-{\rm i}\xi}{2}
 \right )\,
 \left (
 a_{}-{\rm i}b_{}\right )
 \label{redmatrbexbfi}
 \ee
where $T_k(z)$ denotes the $k-$th Tshebyshev polynomial of the
first kind.

One of the latter two relations defines the normalization vector
$(a,b) = (a_0,b_0)$ while their ratio gives
 \be
 T_{n+1}
 \left (
 \frac{-F+{\rm i}\xi}{2}
 \right )\,U_{n+1}
 \left (
 \frac{-F-{\rm i}\xi}{2}
 \right )
+
 T_{n+1}
 \left (
 \frac{-F-{\rm i}\xi}{2}
 \right )\,U_{n+1}
 \left (
 \frac{-F+{\rm i}\xi}{2}
 \right )=0.
 \label{redfinal}
 \ee
This is our final secular equation which defines, in an implicit
manner, the energies $F$ as functions of the couplings~$\xi$.

An efficient numerical treatment of the latter eigenvalue problem
may be based on the re-parametrization
 \be
  \frac{-F+{\rm i}\xi}{2} = \cos \varphi, \ \ \ \ \ \ \ \
  {\rm Re}\,\varphi = \alpha, \ \ \ \ \ \ \
  {\rm Im}\,\varphi = \beta
  \label{mapping}
  \ee
i.e.,
 \be
 \frac{1}{2}\,F=-\cos\alpha \cosh\beta,
 \ \ \ \ \ \ \
 \frac{1}{2}\,\xi=-\sin\alpha \sinh\beta\,.
 \label{dvaro}
 \ee
In opposite direction, the inversion of this change of variables
 \ben
 \cos\alpha=
 -\frac{1}{2\cosh\beta}\,F ,
 \ \ \ \ \ \ \
 \sinh \beta =\frac{1}{2\sqrt{2}}\,
 \sqrt{F^2+\xi^2-4+\sqrt{(F^2+\xi^2-4)^2+16\,\xi^2}}.
 \een
transforms eq.~(\ref{redfinal}) into the compact and transparent
trigonometric secular equation
 \be
  {\rm Re}\,\frac{\sin [(n+1)\varphi]
   \cos [(n+1)\varphi^*]}{\sin \varphi}=0.
   \label{tojevon}
  \ee
Its roots may be determined, numerically, as lying in the domain
with negative $\beta<0$ and with $\alpha \in (0, \pi/2)$ for the
negative $F<0$ and with $\alpha \in (\pi/2,\pi)$ for the positive
$F>0$. In this picture, the constant value of the coupling $\xi>0$
is mapped upon a downwards-oriented half-oval in the
$\alpha-\beta$ plane with a top at $\alpha=\pi/4$. Its two
asymptotes $\alpha= 0$ and $\alpha = \pi/2$ are reached in the
limit $\beta \to -\infty$.

In the new graphical representation the robust, $\xi-$independent
energy level $F=0$ lies on the top of the half-oval while its
decreasing and increasing neighbors are found displaced to the
left and right, respectively, along the half-oval downwards. At
the first few lowest $N=2n+4$ the coordinates of these eigenvalues
remain non-numerical,
 \ben
 \ba
 F_0=0, \ \ F_\pm =\pm \sqrt{2-\xi^2}, \ \ \ \ n = 0,\\
 F_0=0, \ \ F_{\pm,\pm} =\pm \sqrt{2-\xi^2\pm \sqrt{1-4\xi^2}},
  \ \ \ \ \ \ n = 1
 \ea
 \een
etc. The closed form of these definitions enables us to determine
the closed form of the respective critical values at which the
spectrum ceases to be real,
 \ben
 \ba
 Z_{crit}(4)=4\,\sqrt{2} \approx 5.66, \ \ \ \ n = 0,\\
 Z_{crit}(6)=9/{2} =4.50,
  \ \ \ \ \ \ n = 1,
 \ea
 \een
followed by the numerically calculated $Z_{crit}(8) \approx 4.463$
(at $n=2$), $Z_{crit}(10) \approx 4.461$ (at $n=3$), $Z_{crit}(12)
\approx 4.463$ (at $n=4$) etc. These results do not contradict the
expected $n \to \infty$ limit $Z_{crit}(\infty) \approx 4.475$
derived in ref.~\cite{sgezou}.

\subsection{A real-matrix re-arrangement of eq.~(\ref{matr})
}

We may split eq.~(\ref{matr}) in its real and imaginary parts and
`glue' them together in the following very natural pentadiagonal
or, if you wish, block-tridiagonal eigenvalue problem
 \be
 \left (
 \begin{array}{cc|cc|cc|cc|c}
 -F&-\xi&-1&0&&&&&\\
 \xi&-F&0&-1&&&&&\\
 \hline
 -1&0&-F&-\xi&\ddots&&&&\\
 0&-1&\xi&-F&&\ddots&&&\\
 \hline
 &&\ddots&&\ddots&\ddots&-1&0&\\
 &&&\ddots&\ddots&\ddots&0&-1&\\
 \hline
 &&&&-1&0&-F&-\xi&-1\\
 &&&&0&-1&\xi&-F&0\\
 \hline
 &&&&&&-2&0&-F
 \ea
 \right )\,
 \left (
 \ba
 a_0\\
 b_0\\
 \hline
 a_1\\
 b_1\\
 \hline
 \vdots\\
 \vdots \\
 \hline
 a_n\\
 b_n\\
 \hline
 \gamma
 \ea
 \right )
 =0.
 \label{rematr}
 \ee
This equation may be re-written in the partitioned-matrix
notation,
 \be
 \left (
 \begin{array}{ccccc}
 {\rm \bf X}&{\rm \bf -1}&&&\\
 {\rm \bf -1}&{\rm \bf X}&\ddots&&\\
 &\ddots&\ddots&{\rm \bf -1}&\\
 &&{\rm \bf -1}&{\rm \bf X}&\vec{\rm \bf d}\\
 &&&2\vec{\rm \bf d}^{\,T}&-F
 \ea
 \right )\,
 \left (
 \ba
 {\rm \bf \vec{c}_0}\\
 {\rm \bf \vec{c}_1}\\
 \vdots\\
 {\rm \bf \vec{c}_n}\\
 \gamma
 \ea
 \right )
 =0.
 \label{pamatr}
 \ee
The obvious boldface two-by-two matrix elements degenerate, in an
`odd' anomalous last row and column, to an auxiliary vector
$\vec{\rm \bf d}^{\,T}=(1,0)$.

A few comments are due. Firstly, {all} our wave-function
components are now re-interpreted as proportional to the classical
Tshebyshev polynomials $U_k$ with a {two-by-two matrix} argument
$X$. Equation (\ref{pamatr}) leads to an alternative formula for
the eigenvectors,
 \be
 {\rm \bf \vec{c}_k} =U_k
 \left (
 \frac{1}{2}
 {\rm \bf X}
 \right )\,
 {\rm \bf \vec{c}_0}\,,
 \ \ \ \ \ \ \
 {\rm \bf X}=
 \left (
 \begin{array}{cc}
 -F&-\xi\\
 \xi&-F
 \ea
 \right )\,,
 \ \ \ \ \ {\rm \bf k} =0,1,\ldots,n+1.
 \label{closedf}
 \ee
Secondly, once we introduce a {\em complex} angle $\alpha$ we may
parametrize $F=-\varrho\,\cos \alpha$ and $\xi = \varrho \,\sin
\alpha$ {using an optional}, redundant parameter $\varrho$. A
peculiar feature of our matrices $X$ is that their powers remain
elementary in this representation,
 \ben
 {\rm \bf X}^m =\varrho^m\,
 \left (
 \begin{array}{cc}
 \cos m \alpha&- \sin m \alpha\\
  \sin m \alpha& \cos m \alpha
 \ea
 \right )\,.
 \een
This means that all the formulae containing polynomials
(\ref{closedf}) remain amazingly transparent.

The existence of the explicit solutions (\ref{closedf}) reduces
eq.~(\ref{pamatr}) to the two secular-equation constraints imposed
upon the vector ${\rm \bf \vec{c}_{n+1}}$. Of course, they are
equivalent to our complex matching conditions as mentioned above.

\section{Solutions at the odd $N = 2n+3$}

Weigert \cite{Weigert} did not notice that a ``one-step easier"
discretization of eq.~(\ref{basic}) emerges at the odd $N = 2n+3$,
with $2N+2$ energy roots at $n\geq 0$. An alternative to
eq.~(\ref{matr}) then reads, in the same notation,
 \be
 \left (
 \begin{array}{cccc|ccc}
 {\rm i}\xi-F&-1&&&&&\\
 -1&{\rm i}\xi-F&\ddots&&&&\\
 &\ddots&\ddots&-1&&&\\
 &&-1&{\rm i}\xi-F&-1&&\\
 \hline
 &&&-1&-{\rm i}\xi-F&\ddots&\\
 &&&&\ddots&\ddots&-1\\
 &&&&&-1&-{\rm i}\xi-F
 \ea
 \right )\,
 \left (
 \ba
 \alpha_0\\
 \alpha_1\\
 \vdots\\
 \alpha_n\\
 \hline
 \alpha^*_n\\
 \vdots\\
 \alpha^*_0
 \ea
 \right )
 =0.
 \label{patr}
 \ee
Definition (\ref{eigenvecs}) of the eigenvectors remains unchanged
but the matching condition is just one,
 \be
 \gamma= U_{n+1}
 \left (
 \frac{-F+{\rm i}\xi}{2}
 \right )\,
 \left (
 a_{}+{\rm i}b_{}\right )
 =
 U_{n}
 \left (
 \frac{-F-{\rm i}\xi}{2}
 \right )\,
 \left (
 a_{}-{\rm i}b_{}\right ).
 \label{redmatrbeno}
 \ee
The ratio between this equation and its Hermitian conjugate
eliminates all the normalization ambiguities and leads to the
odd$-N$ counterpart of eq.~(\ref{redfinal}),
 \be
 U_{n}
 \left (
 \frac{-F+{\rm i}\xi}{2}
 \right )\,U_{n}
 \left (
 \frac{-F-{\rm i}\xi}{2}
 \right )
 =U_{n+1}
 \left (
 \frac{-F+{\rm i}\xi}{2}
 \right )\,
 U_{n+1}
 \left (
 \frac{-F-{\rm i}\xi}{2}
 \right )\,.
 \label{redfinalbe}
 \ee
This secular equation is our final result. As an implicit
definition of the $N=2n+3$ energy levels $F = F(\xi)$ it possesses
the compact non-numerical solutions at the first two values of $n$
again,
 \ben
 \ba
  F_\pm =\pm \sqrt{1-\xi^2}, \ \ \ \ n = 0,\\
  F_{\pm,\pm} =\pm \frac{1}{2}\,\sqrt{6-4\xi^2\pm 2\,\sqrt{5-16\xi^2}},
  \ \ \ \ \ \ n = 1.
 \ea
 \een
The respective elementary expressions for the critical constants
 \ben
 \ba
 Z_{crit}(3)=\frac{9}{4}=2.25, \ \ \ \ n = 0,\\
 Z_{crit}(5)=\frac{25\,\sqrt{5}}{16} \approx 3.49,
  \ \ \ \ \ \ n = 1
 \ea
 \een
are followed by the complex Cardano representation of the real
$Z_{crit}(7) \approx 3.946$ at $n=2$. At $n > 2$ one switches to a
purely numerical algorithm giving $Z_{crit}(9) \approx 4.148$ at
$n=3$ etc. In comparison with the parallel results sampled in
section~\ref{rabove} at even~$N$ we notice a slowdown of the
numerical convergence towards the $n \to \infty$ limit.

\subsection*{Acknowledgment}

Proportionally supported by NPI, IRP AV0Z10480505, and by GA AS,
contract No. A 1048302.

\end{document}